\documentclass[journal,onecolumn]{IEEEtran}

\usepackage{balance} 
\usepackage{hyperref}       
\usepackage{url}            
\usepackage{booktabs}       
\usepackage{nicefrac}       
\usepackage{microtype}      
\usepackage{subcaption}
\usepackage{graphicx}
\usepackage{amsmath,mathtools,amsfonts,amsthm} 
\usepackage{xcolor}
\usepackage{subcaption}

\begin{document}


\title{Non-Parametric Neuro-Adaptive Coordination of Multi-Agent Systems}

\author{Christos~K.~Verginis,
	Zhe~Xu,
	and~Ufuk~Topcu
	\thanks{C. K. Verginis and U. Topcu are with the University of Texas at Austin,
		Austin, Texas, USA, e-mail: \{cverginis, utocpu\}@utexas.edu. }
	\thanks{Z. Xu is with Arizona State University, Tempe, Arizona, USA, e-mail: xzhe1@asu.edu}
}

\maketitle
\begin{abstract}
We develop a learning-based algorithm for the distributed formation control
of networked multi-agent systems governed by unknown, nonlinear dynamics. 
The proposed algorithm integrates neural network-based learning with adaptive control in a two-step procedure. 
In the first step, each agent learns a controller, represented as a neural network, using training data that correspond to a collection of formation tasks and agent parameters. 
These parameters and tasks are derived by varying the nominal agent parameters and the formation specifications of the task in hand, respectively.
In the second step of the algorithm, each agent incorporates the trained neural network into an online and adaptive control policy in such a way that the behavior of the multi-agent closed-loop system satisfies a user-defined formation task. Both the learning phase and the adaptive control policy are distributed, in the sense that each agent computes its own actions using only local information from its neighboring agents. 
\end{abstract}









\section{Introduction}

During the last decades, decentralized control of networked multi-agent systems has attracted significant attention due to the great variety of its applications, including multi-robot systems, transportation, multi-point surveillance as well as biological systems \cite{jadbabaie2003coordination,olfati2007consensus,couzin2005effective,modares2017optimal,bechlioulis2016decentralized,verginis2019adaptive,ni2020adaptive,zhang2012adaptive,hu2014adaptive}. 
{In such systems, each agent calculates its own actions based on local information, as modeled by a connectivity graph, without relying on any central control unit. 

Although many works on distributed cooperative control consider known and simple dynamic models, there exist many practical engineering systems that cannot be modeled accurately and 
are affected by unknown exogenous disturbances. Thus, the design of control algorithms that are robust and adaptable to such uncertainties and disturbances is important. For multi-agent systems, ensuring robustness is particularly challenging due to the lack of global information and the interacting dynamics of the individual agents. {A promising step towards the control of systems with uncertain dynamics is the use of data obtained a priori from system runs. However, engineering systems often undergo purposeful modifications (e.g., substitution of a motor or link in a robotic arm or exposure to new working environments) or suffer gradual faults (e.g., mechanical degradation), which might change the systems' dynamics or operating conditions. Therefore, one cannot rely on the aforementioned data to provably guarantee the successful control of the system. 
On the other hand, the exact incorporation of these changes in the dynamic model, 
and consequently, the design of new model-based algorithms, can be a challenging procedure. Hence, the goal in such cases is to exploit the data obtained a priori and construct intelligent online policies that achieve a user-defined task while
adapting to the aforementioned changes. 
 }


{
This paper addresses the distributed coordination of networked multi-agent systems governed by unknown nonlinear dynamics. 
We design a control algorithm that draws a novel connection between distributed learning with neural-network-based representations and adaptive feedback control, and consists of the following steps. 
Firstly, it trains a number of neural networks, one for each agent, to approximate controllers for the agents that accomplish the given formation task. The data used to train the neural networks consist of pairs of states and control actions of the agents that are gathered from runs of the multi-agent system. 
Secondly, it uses an online adaptive feedback control policy that guarantees accomplishment of the given formation task. 
Both steps can be executed in a distributed manner in a sense that each agent uses only local information, as modeled by a connectivity graph. 
}

\section{Control Algorithm} \label{sec:PF}

Consider a networked multi-agent group comprised of a leader, indexed by $i=0$, and $N$ followers, with $\mathcal{N}\coloneqq\{1,\dots,N\}$. The leading agent acts as an exosystem that generates a desired reference trajectory for the multi-agent group. The followers, which have to be controlled,
evolve according to the $2$nd-order dynamics
\begin{subequations} \label{eq:dynamics}	
\begin{align}
	\dot{x}_{i,1}(t) &= x_{i,2}(t) \\
	\dot{x}_{i,2}(t) &= f_i(x_i(t),t) + g_i({x}_i(t),t)u_i(t)
\end{align}
\end{subequations}
where ${x}_i \coloneqq [x_{i,1}^\top, x_{i,2}^\top]^\top \in \mathbb{R}^{2n}$ is the $i$th agent's state, assumed available for measurement by agent $i$, $f_i:\mathbb{R}^{2n}\times[0,\infty) \to \mathbb{R}^n$, $g_i:\mathbb{R}^{2n}\times[0,\infty) \to \mathbb{R}^n$ are unknown functions modeling the agent's dynamics, and $u_i$ is the $i$th agent's control input. The vector fields $f_i(\cdot)$ and $g_i(\cdot)$ are assumed to be locally Lipschitz in ${x}_i$ over $\mathbb{R}^{2n}$ for each fixed $t\geq 0$, and uniformly bounded in $t$ over $[t_0,\infty)$ for each fixed ${x}_i\in\mathbb{R}^{2n}$, for all $i\in\mathcal{N}$. Further, we assume that the matrices $g_i$ are positive definite, for all $i\in\mathcal{N}$. 

We use an undirected graph $\mathcal{G} \coloneqq (\mathcal{N},\mathcal{E})$ to model the communication among the agents, with $\mathcal{N}$ being the index set of the agents. 
The set of neighbors of agent $i$ is denoted by $\mathcal{N}_i \coloneqq \{j\in\mathcal{N}:(i,j)\in\mathcal{E}\}$. 
We assume that $\mathcal{G}$ is connected, i.e., 
there exists a communication path between any two agents. 

The state/command variables of the leading agent (indexed by $0$) are denoted by $x_{0,1}$, $x_{0,2}$ $\in\mathbb{R}^n$ and
obey the $2$nd-order dynamics $\dot{x}_{0,1}(t) = x_{0,2}(t)$, $\dot{x}_{0,2}(t) = u_{0}(t)$ 
for a smooth and bounded $u_0:[0,\infty) \to \mathbb{R}^n$. However, the state of the leader is only provided to a subgroup of the $N$ agents. In particular, the access of the follower agents to the leader's state is modeled by a diagonal matrix $\mathcal{B} \coloneqq \textup{diag}\{b_1,\dots,b_N\} \in \mathbb{R}^{N\times N}$; if $b_i = 1$, then the $i$th agent has access to the leader's state, whereas it does not if $b_i = 0$, for $i\in\mathcal{N}$. 
Thus, we may also define the augmented graph as $\bar{\mathcal{G}} \coloneqq (\mathcal{N}\cup\{0\}, \bar{\mathcal{E}})$, where $\bar{\mathcal{E}} \coloneqq \mathcal{E} \cup \{ (i,0) : b_i = 1 \}$.   
The goal of this work is to design a distributed control algorithm, where each agent has access only to its neighbors' information, to achieve a pre-specified geometric formation of the agents in $\mathbb{R}^n$. More specifically, consider for each agent $i\in\mathcal{N}$ the constants $c_{ij}$, $j\in \{0\} \cup \mathcal{N}_i$ prescribing a desired offset that agent $i$ desires to achieve with respect to the leader ($j=0$), and its neighbors ($j\in\mathcal{N}_i$). That is, each agent $i\in\mathcal{N}_i$ aims at achieving $x_{i,1} = x_{j,1} - c_{ij}$, for all $j\in\mathcal{N}_i$, and if $b_i=1$, $x_{i,1} = x_{0,1} - c_{i0}$. 
In other words, we aim to minimize the errors, 
\begin{align*} 
	e_{i,1} \coloneqq \sum_{j\in\mathcal{N}_i} (x_{i,1} - x_{j,1} + c_{ij}) + b_i(x_{i,1} - x_{0,1} + c_{i0}), \ \ \ i\in\mathcal{N}.
\end{align*}
We describe now the control algorithm. 
We assume the existence of data gathered from a finite set of $T$ trajectories $\mathcal{J}$ generated by a priori runs of the multi-agent system. 
More specifically, we consider that $\mathcal{J}$ is decomposed as $\mathcal{J} = (\mathcal{J}_1,\dots,\mathcal{J}_N)$, where $\mathcal{J}_i$ is the set of trajectories 
$$\mathcal{J}_i = \left\{\bar{x}^k_i(t), \{\bar{x}^j\}_{j\in\mathcal{N}^k_i},u^k_i \left(\bar{x}^k_i(t), \{\bar{x}^j\}_{j\in\mathcal{N}^k_i},t \right) \right\}_{t\in \mathbb{T}_i}$$
of agent $i$,
where $\mathbb{T}_i$ is a finite set of time instants, $\bar{x}^k_i\in\mathbb{R}^{2n}$ is the state trajectory of agent $i$ for trajectory $k$, $\mathcal{N}^k_i$ are the neighbors of agent $i$ in trajectory $k$, with $\{\bar{x}^j\}_{j\in\mathcal{N}^k_i}$ being their respective state trajectories, and $u^k_i(\bar{x}^k_i(t), \{\bar{x}^j\}_{j\in\mathcal{N}^k_i},t) \in \mathbb{R}^n$ is the  control input trajectory of agent $i$. 

Each agent $i\in\mathcal{N}$ uses the data to train a neural network in order to approximate a controller that accomplishes the formation task. 
More specifically, each agent uses the tuples  $\{\bar{x}^k_i(t), \{\bar{x}^j\}_{j\in\mathcal{N}^k_i}\}_{t\in\mathbb{T}_i}$ as input to a neural network, and $u^k_i \big(\bar{x}^k_i(t), \{\bar{x}^j\}_{j\in\mathcal{N}^k_i},t \big)_{t\in\mathbb{T}_i}$ as the respective output targets, for all $T$ trajectories. 
For the inputs corresponding to agents that are not neighbors of agent $i$ in a trajectory $k$, we disable the respective neurons. 
For a given $\bar{x} \in \mathbb{R}^{2n}$, we denote by $u_{i,nn}(\bar{x})$ the output of the neural network of agent $ i \in \mathcal{N}$.

We now design a distributed, adaptive feedback control policy to accomplish the formation task.
Consider the adaptation variables $\hat{d}_{i,1}$ and $\hat{d}_{i,2}$ for each agent $i\in\mathcal{N}$, corresponding to upper bounds of the unknown dynamic terms $f_i$ and $g_i$. 
Consider  the augmented errors for each agent $e_{i,2} \coloneqq \dot{e}_{i,1} + k_{i,1}e_{i,1}$,
 where $k_{i,1}$ are positive constants, for all $i\in\mathcal{N}$. We design the distributed control policy as 
\begin{subequations} \label{eq:control and adapt}
\begin{align} 
	u_i(\bar{x},\hat{d}_{i,1},\hat{d}_{i,2}) = u_{i,nn}(\bar{x}) -(k_{i,2}  + \hat{d}_{i,1})e_{i,2} -  \hat{d}_{i,2} \hat{e}_{i,2}
\end{align}
where $k_{i,2}$ are positive constants, and $\hat{e}_{i,2}$ are defined as $\hat{e}_{i,2} \coloneqq \frac{e_{i,2}}{\|e_{i,2}\|^2}$ if $e_{i,2} \neq 0$ and $\hat{e}_{i,2} \coloneqq 0$ otherwise, for all $i\in\mathcal{N}$. The adaptation variables $\hat{d}_{i,1}$, $\hat{d}_{i,2}$ are updated as 
\begin{align*} \label{eq:adaptation law}
		\dot{\hat{d}}_{i,1} \coloneqq \mu_{i,1}\|e_{i,2}\|^2, \hspace{5mm}
		\dot{\hat{d}}_{i,2} \coloneqq \mu_{i,2}\|e_{i,2}\|,
\end{align*}
\end{subequations}
where $\mu_{i,1}$, $\mu_{i,2}$ are positive constants, for all $i\in\mathcal{N}$.


\begin{figure}
	\centering
	\includegraphics[width=.35\textwidth]{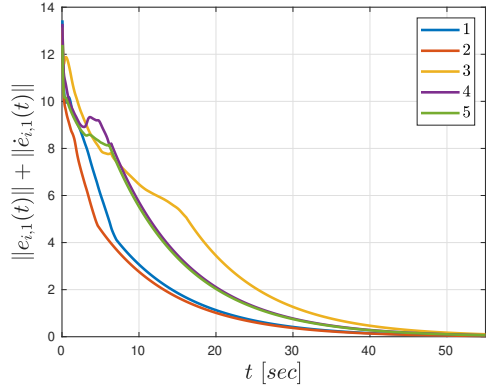}
	\caption{Evolution of the error signals $\|e_{i,1}(t)\|+\|\dot{e}_{i,1}(t)\|$ for $i\in\{1,\dots,5\}$, and $t\in[0,55]$, for the numerical experiments.}
	\label{fig:AS errors}
\end{figure}

\begin{figure}
	\centering
	\includegraphics[width=.3\textwidth]{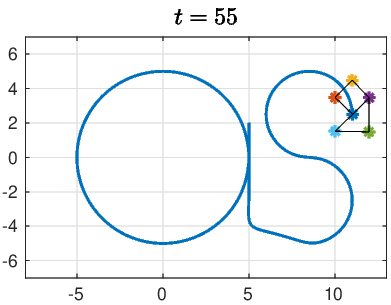}
	\caption{The convergence of the followers to the desired formation around the leader, which follows a pre-specified trajectory (continuous blue line), in the $x$-$y$ plane. 
	}
	\label{fig:ext_plot}
\end{figure}

\section{Numerical Experiments} \label{sec:exps}
We consider $N=5$ follower aerial vehicles in $\mathbb{R}^3$ with dynamics of the form \eqref{eq:dynamics}, with communication graph modeled by the edge set $\bar{\mathcal{E}}$ $=$ $\{$ $(1,2)$, $(2,3)$, $(3,4)$, $(4,5)$, $(1,0)$, $(3,0)$, $(5,0)$ $\}$.
The leader's task is to track a reference time-varying trajectory profile $x_0(t)$. 
The formation constants $c_{ij}$ are chosen randomly in $(-1,1)$, $(i,j)\in\bar{\mathcal{E}}$. 
We generate data from $100$ trajectories that correspond to different $f_i$, ${g}_i$, and initial conditions,
and we train $5$ neural networks, one for each agent. We test the control policy \eqref{eq:control and adapt} and obtain 
the results depicted in Fig \ref{fig:AS errors}, which shows the evolution of the error signals $\|e_{i,1}(t)\| + \|\dot{e}_{i,1}(t)\|$ for $i\in\{1,\dots,5\}$. One concludes that the multi-agent system converges successfully to the pre-specified formation since the error signals converge to zero.

\bibliographystyle{IEEEtran} 
\bibliography{sample}

\end{document}